\documentclass[sigconf]{acmart}

\AtBeginDocument{%
  \providecommand\BibTeX{{%
    \normalfont B\kern-0.5em{\scshape i\kern-0.25em b}\kern-0.8em\TeX}}}

\setcopyright{acmcopyright}
\copyrightyear{2019}
\acmYear{2019}
\acmDOI{10.1145/1122445.1122456}

\acmConference[SoCC '19]{ACM Symposium of Cloud Computing conference}{November 20-23}{Santa Cruz, CA}
\acmBooktitle{SoCC '19: ACM Symposium of Cloud Computing conference,
  Nov 20--23, 2019, Santa Cruz, CA}
\acmPrice{15.00}
\acmISBN{978-1-4503-9999-9/18/06}

\begin{document}

%%
%% The "title" command has an optional parameter,
%% allowing the author to define a "short title" to be used in page headers.
\title{BigDL: A Distributed Deep Learning Framework for Big Data }

%%
%% The "author" command and its associated commands are used to define
%% the authors and their affiliations.
%% Of note is the shared affiliation of the first two authors, and the
%% "authornote" and "authornotemark" commands
%% used to denote shared contribution to the research.
\author{Jason (Jinquan) Dai}
\affiliation{\institution{Intel Corporation}}

\author{Yiheng Wang}
\authornote{Work was done when the author worked at Intel}
\affiliation{\institution{Tencent Inc.}}

\author{Xin Qiu}
\affiliation{\institution{Intel Corporation}}

\author{Ding Ding}
\affiliation{\institution{Intel Corporation}}

\author{Yao Zhang}
\authornotemark[1]
\affiliation{\institution{Sequoia Capital}}

\author{Yanzhang Wang}
\affiliation{\institution{Intel Corporation}}

\author{Xianyan Jia}
\authornotemark[1]
\affiliation{\institution{Alibaba Group}}

\author{Cherry (Li) Zhang}
\affiliation{\institution{Intel Corporation}}

\author{Yan Wan}
\authornotemark[1]
\affiliation{\institution{Alibaba Group}}

\author{Zhichao Li}
\affiliation{\institution{Intel Corporation}}

\author{Jiao Wang}
\affiliation{\institution{Intel Corporation}}

\author{Shengsheng Huang}
\affiliation{\institution{Intel Corporation}}

\author{Zhongyuan Wu}
\affiliation{\institution{Intel Corporation}}

\author{Yang Wang}
\affiliation{\institution{Intel Corporation}}

\author{Yuhao Yang}
\affiliation{\institution{Intel Corporation}}

\author{Bowen She}
\affiliation{\institution{Intel Corporation}}

\author{Dongjie Shi}
\affiliation{\institution{Intel Corporation}}

\author{Qi Lu}
\affiliation{\institution{Intel Corporation}}

\author{Kai Huang}
\affiliation{\institution{Intel Corporation}}

\author{Guoqiong Song}
\affiliation{\institution{Intel Corporation}}

%%
%% By default, the full list of authors will be used in the page
%% headers. Often, this list is too long, and will overlap
%% other information printed in the page headers. This command allows
%% the author to define a more concise list
%% of authors' names for this purpose.
\renewcommand{\shortauthors}{J. Dai et al.}

%%
%% The abstract is a short summary of the work to be presented in the
%% article.
\begin{abstract}
  This paper presents BigDL (a distributed deep learning framework for Apache Spark), which has been used by a variety of users in the industry for building deep learning applications on production big data platforms. It allows deep learning applications to run on the Apache Hadoop/Spark cluster so as to directly process the production data, and as a part of the end-to-end data analysis pipeline for deployment and management. Unlike existing deep learning frameworks, BigDL implements distributed, data parallel training directly on top of the functional compute model (with copy-on-write and coarse-grained operations) of Spark. We also share real-world experience and "war stories" of users that have adopted BigDL to address their challenges(i.e., how to easily build end-to-end data analysis and deep learning pipelines for their production data).
\end{abstract}

%%
%% The code below is generated by the tool at http://dl.acm.org/ccs.cfm.
%% Please copy and paste the code instead of the example below.
%%


%%
%% Keywords. The author(s) should pick words that accurately describe
%% the work being presented. Separate the keywords with commas.

%%
%% This command processes the author and affiliation and title
%% information and builds the first part of the formatted document.
\maketitle

\begin{figure*}
\centering
\includegraphics[width=0.95\linewidth]{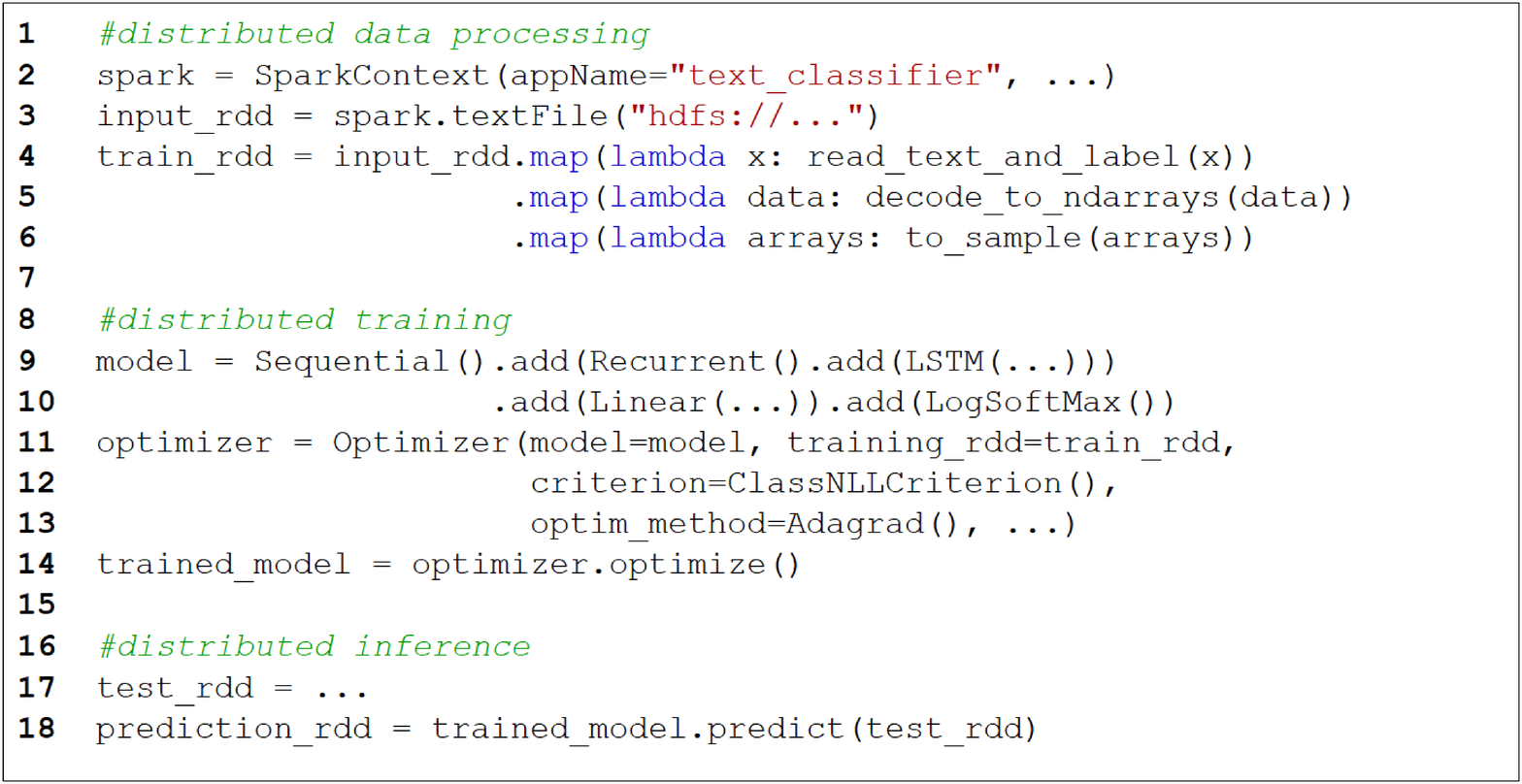}
\caption{The end-to-end text classification pipeline (including data loading, processing, training, prediction, etc.) on Spark and BigDL}
\label{fig:fig1}
\end{figure*}

\section{Introduction}
Continued advancements in artificial intelligence applications have brought deep learning to the forefront of a new generation of data analytics development; as the requirements and usage models expand, new systems and architecture beyond existing deep learning frameworks (e.g., Caffe  \cite{jia2014caffe}, Torch \cite{collobert2011torch7}, TensorFlow  \cite{abadi2016tensorflow}, MXNet \cite{chen2015mxnet}, Chainer \cite{tokui2015chainer}, PyTorch \cite{paszke2017automatic}, etc.) have inevitably emerged. In particular, there is increasing demand from organizations to apply deep learning technologies to their big data analysis pipelines.

To support these new requirements, we have developed BigDL, a distributed deep learning framework for big data platforms and workflows. It is implemented as a library on top of Apache Spark \cite{spark}, and allows users to write their deep learning applications as standard Spark programs, running directly on existing big data (Apache Hadoop \cite{hadoop} or Spark) clusters. It supports an API similar to Torch and Keras \cite{keras} for constructing neural network models (as illustrate in Figure 1); it also supports both large-scale distributed training and inference, leveraging the scale-out architecture of the underlying Spark framework (which runs across hundreds or thousands of servers efficiently).

BigDL provides an expressive, "data-analytics integrated" deep learning programming model; within a single, unified data analysis pipeline, users can efficiently process very large dataset using Spark APIs (e.g., RDD \cite{zaharia2012resilient}, Dataframe \cite{armbrust2015spark}, Spark SQL, ML pipeline, etc.), feed the distributed dataset to the neural network model, and perform distributed training or inference on top of Spark. Contrary to the conventional wisdom of the machine learning community (that fine-grained data access and in-place updates are critical for efficient distributed training \cite{abadi2016tensorflow}), BigDL provides large-scale, data parallel training directly on top of the functional compute model (with copy-on-write and coarse-grained operations) of Spark. By unifying the execution model of neural network models and big data analytics, BigDL allows new deep learning algorithms to be seamless integrated into production data pipelines, which can then be easily deployed, monitored and managed in a single unified big data platform.

BigDL is developed as an open source project\footnote{https://github.com/intel-analytics/BigDL}; over the past years, a variety of users in the industry (e.g., Mastercard, World Bank, Cray, Talroo, UCSF, JD, UnionPay, Telefonica, GigaSpaces, etc.) have built their data analytics and deep learning applications on top of BigDL for a wide range of workloads, such as transfer learning based image classification, object detection and feature extraction, sequence-to-sequence prediction for precipitation nowcasting, neural collaborative filtering for recommendations, etc. In this paper, we focus on the execution model of BigDL to support large-scale distributed training (a challenging system problem for deep learning frameworks), as well as empirical results of real-world deep learning applications built on top of BigDL. The main contributions of this paper are:

\begin{itemize}
\item It presents BigDL, a working system that have been used by many users in the industry for distributed deep learning on production big data systems.
\item It describes the distributed execution model in BigDL (that adopts the state of practice of big data systems), which provides a viable design alternative for distributed model training (compared to existing deep learning frameworks).
\item It shares real-world experience and "war stories" of users that have adopted BigDL to address their challenges (i.e., how to easily build end-to-end data analysis and deep learning pipelines for their production data).
\end{itemize}

\section{Motivation}

A lot of efforts in the deep learning community have been focusing on improving the accuracy and/or speed of standard deep learning benchmarks (such as ImageNet \cite{russakovsky2015imagenet} or SQuAD \cite{rajpurkar2016squad}). For these benchmarks, the input dataset have already been curated and explicitly labelled, and it makes sense to run deep learning algorithms on specialized deep learning frameworks for best computing efficiency. On the other hand, if the input dataset are dynamic and messy (e.g., live data streaming into production data pipeline that require complex processing), it makes more sense to adopt BigDL to build the end-to-end, integrated data analytics and deep learning pipelines for production data.

As mentioned in Section 1, BigDL has been used by a variety of users in the industry to build deep learning applications on their production data platform. The key motivation for adopting such a unified data analytics and deep learning system like BigDL is to improve the ease of use (including development, deployment and operations) for applying deep learning in real-world data pipelines.

In real world, it is critical to run deep learning applications directly on where the data are stored, and as a part of the end-to-end data analysis pipelines. Applying deep learning to production big data is very different from the ImageNet \cite{russakovsky2015imagenet} or SQuAD \cite{rajpurkar2016squad} problem; real-world big data are both dynamic and messy, and are possibly implicitly labeled (e.g., implicit feedbacks in recommendation applications \cite{jawaheer2010comparison}), which require very complex data processing; furthermore, instead of running ETL (extract, transform and load) and data processing only once, real-world data analytics pipeline is an iterative and recurrent process (e.g., back-and-forth development and debugging, incremental model update with new production data, etc.). Therefore, it is highly inefficient to run these workloads on separate big data and deep learning systems (e.g., processing data on a Spark cluster, and then export the processed data to a separate TensorFlow cluster for training/inference) in terms of not only data transfer, but also development, debugging, deployment and operation productivity.

One way to address the above challenge is to adopt a "connector approach" (e.g., TFX \cite{baylor2017tfx}, CaffeOnSpark \cite{Yahoo2016}, TensorFlowOnSpark \cite{Yahoo2017}, SageMaker \cite{Amazon2017}, etc.), which develops proper interfaces to connect different data processing and deep learning components using an integrated workflow (and possibly on a shared cluster). However, the adaptation between different frameworks can impose very large overheads in practice (e.g., inter-process communication, data serialization and persistency, etc.). More importantly, this approach suffers from impedance mismatches \cite{lin2013scaling} that arise from crossing boundaries between heterogeneous components. For instance, many of these systems (such as TensorFlowOnSpark) first use big data (e.g., Spark) tasks to allocate resources (e.g., Spark worker nodes), and then run deep learning (e.g., TensorFlow) tasks on the allocated resources. However, big data and deep learning systems have very different distributed execution model - big data tasks are embarrassingly parallel and independent of each other, while deep learning tasks need to coordinate with and depend on others. For instance, when a Spark worker fails, the Spark system just relaunch the worker (which in turn re-runs the TensorFlow task); this however is incompatible with the TensorFlow execution model and can cause the entire workflow to block indefinitely.

The Big Data community have also started to provide better support for the "connector approach". For instance, the barrier execution mode introduced by Project Hydrogen \cite{ReynoldXin2018} provides gang scheduling \cite{GangScheduling} support in Spark, so as to overcome the errors caused by different execution models between Spark and existing deep learning frameworks (as described in the preceding paragraph). On the other hand, this does not eliminate the difference in the two execution models, which can still lead to lower efficiency (e.g., it is unclear how to apply delay scheduling \cite{zaharia2010delay} to gang scheduling in Spark, resulting in poorer data locality). In addition, it does not address other impedance mismatches such as different parallelism behaviors between data processing and model computations (e.g., see Section 5.1).

BigDL has taken a different approach that directly implements the distributed deep learning support in the big data system (namely, Apache Spark). Consequently, one can easily build the end-to-end, "data-analytics integrated" deep learning pipelines (under a unified programming paradigm, as illustrated in Figure 1), which can then run as standard Spark jobs to apply large-scale data processing and deep learning training/inference to production dataset within a single framework. This completely eliminates the impedance mismatch problems, and greatly improves the efficiency of development and operations of deep learning applications for big data.

\section{BigDL Execution Model}

This section describes in detail how BigDL support large-scale, distributed training on top of Apache Spark. While it has adopted the standard practice (such as data parallel training \cite{dean2012large}, parameter server and AllReduce \cite{abadi2016tensorflow}  \cite{li2014scaling} \cite{chilimbi2014project} \cite{xing2015petuum}) 
\cite{zhang2017poseidon}for scalable training, the key novelty of BigDL is how to efficiently implement these functionalities on a functional, coarse-grained compute model of Spark.

The conventional wisdom of the machine learning community is that, fine-grained data access and in-place data mutation are critical to support highly efficient parameter server, AllReduce and distributed training \cite{abadi2016tensorflow}. On the other hand, big data systems (such as Spark) usually adopts a very different, functional compute model, where dataset are immutable and can only be transformed into new dataset without side effects (i.e., copy-on-write); in addition, the transformations are coarse-grained operations (i.e., applying the same operation to all data items at once).

\begin{figure}[ht]
\centering
\includegraphics[width=1\linewidth]{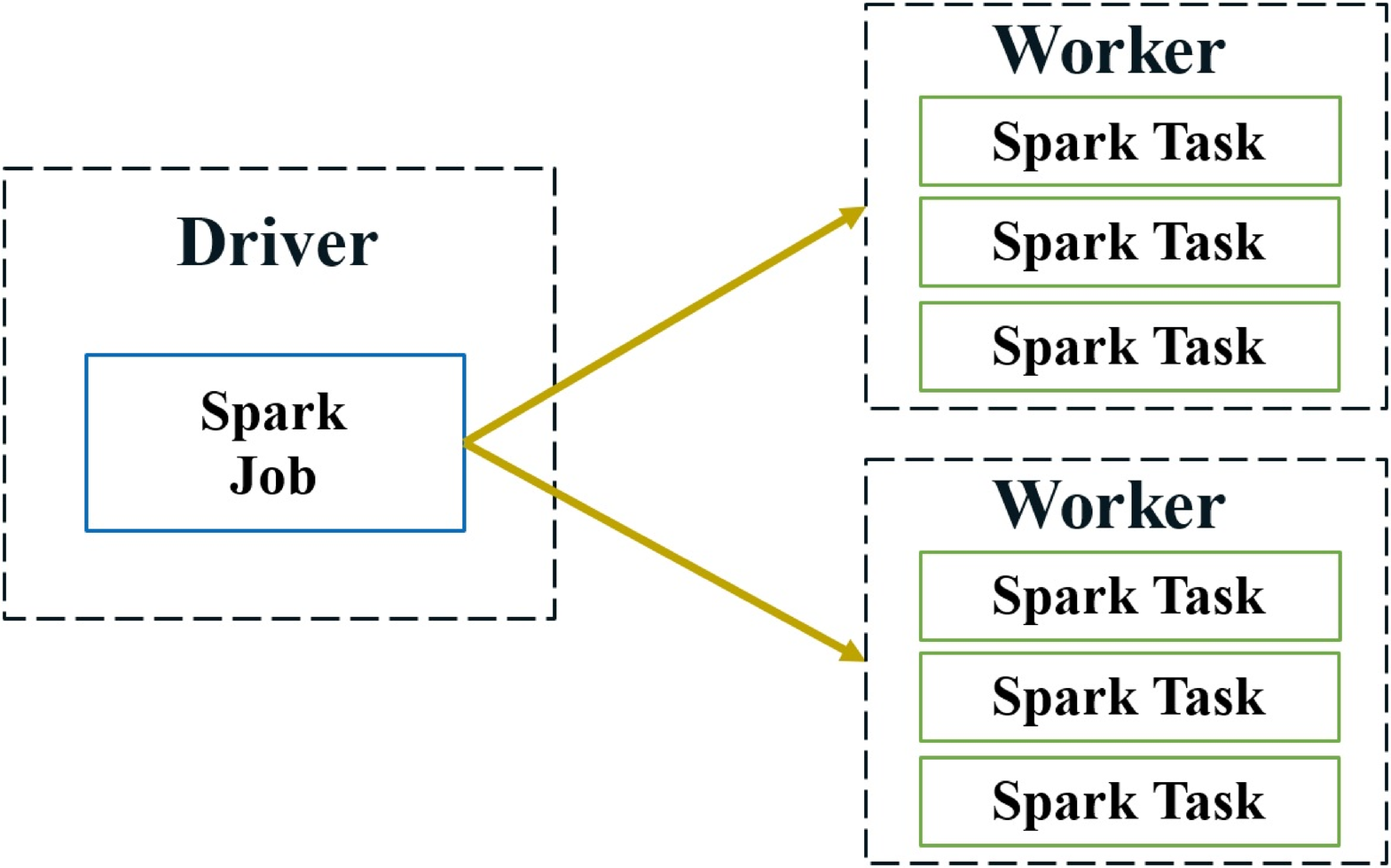}
\caption{A Spark job consists of many Spark tasks; the driver node is responsible for scheduling and dispatching the tasks to worker nodes, which runs the actual Spark tasks.}
\label{fig:fig2}
\end{figure}

\begin{table}[ht]
\centering
\begin{tabular}{l}
\toprule
\emph{Algorithm 1 Data-parallel training in BigDL} \\
\midrule
1: \textbf{for} \emph{i} = 1 to \emph{M} \textbf{do} \\
2:\qquad//"model forward-backward" job \\
3:\qquad\textbf{for} each task in the Spark job \textbf{do} \\
4:\qquad \quad read the latest \textbf{weights}; \\
5:\qquad \quad get a random \textbf{batch} of data from local \emph{Sample} partition; \\
6:\qquad \quad compute local \textbf{gradients} (forward-backward
 on local \emph{model} \\
\qquad \quad \, \, replica); \\
7:\qquad \textbf{end for} \\
8:\qquad //"parameter synchronization" job \\
9:\qquad aggregate (sum) all the \textbf{gradients}; \\
10:\quad \, update the \textbf{weights} per specified optimization method; \\
11:  \textbf{end for} \\
\bottomrule
\end{tabular}
\end{table}

BigDL is implemented as a standard library on Spark and has adopted this functional compute model; nevertheless, it still provides an efficient "parameter server" style architecture for efficient distributed training (by implementing an AllReduce like operation directly using existing primitives in Spark).

\begin{figure*}
\centering
\includegraphics[width=1\linewidth]{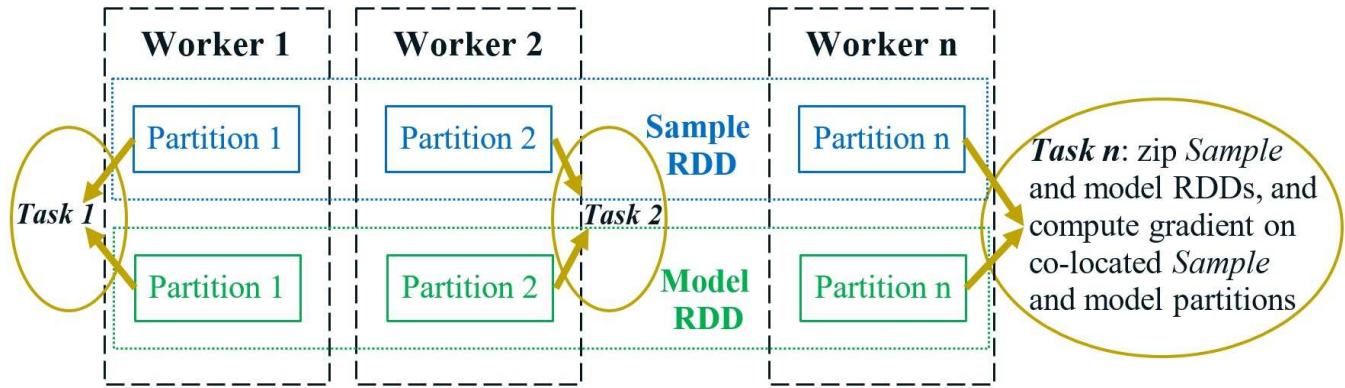}
\caption{The "model forward-backward" spark job, which computes the local gradients for each model replica in parallel.}
\label{fig:fig3}
\end{figure*}

\subsection{Spark execution model}

Similar to other Big Data systems (such as MapReduce \cite{dean2008mapreduce} and Dryad \cite{isard2007dryad}), a Spark cluster consists of a single driver node and multiple worker nodes, as shown in Figure 2. The driver is responsible for coordinating tasks in a Spark job (e.g., task scheduling and dispatching), while the workers are responsible for the actual computation. To automatically parallelize the data processing across the cluster in a fault-tolerant fashion, Spark provides a data-parallel, functional compute model. In a Spark job, data are represented as Resilient Distributed Dataset (RDD) \cite{zaharia2012resilient}, which is an immutable collection of records partitioned across a cluster, and can only be transformed to derive new RDDs (i.e., copy-on-write) through functional operators like \emph{map}, \emph{filter} and \emph{reduce} (e.g., see line 4 - 6 in Figure 1); in addition, these operations are both data-parallel (i.e., applied to individual data partitions in parallel by different Spark tasks) and coarse-grained (i.e., applying the same operation to all data items at once).

\subsection{Data-parallel training in BigDL}

Built on top of the data-parallel, functional compute model of Spark, BigDL provides synchronous data-parallel training to train a deep neural network model across the cluster, which is shown to achieve better scalability and efficiency (in terms of time-to-quality) compared to asynchronous training \cite{chen2016revisiting}. Specifically, the distributed training in BigDL is implemented as an iterative process, as illustrated in Algorithm 1; each iteration runs a couple of Spark jobs to first compute the gradients using the current mini-batch (by a "model forward-backward" job), and then make a single update to the parameters of the neural network model (by a "parameter synchronization" job).

The training data in BigDL are represented as an RDD of Samples (see line 6 in Figure 1), which are automatically partitioned across the Spark cluster. In addition, to implement the data-parallel training, BigDL also constructs an RDD of models, each of which is a replica of the original neural network model. Before the training, both the model and Sample RDDs are cached in memory, and co-partitioned and co-located across the cluster, as shown in Figure 3; consequently, in each iteration of the model training, a single "model forward-backward" Spark job can simply apply the functional \emph{zip} operator to the co-located partitions of the two RDDs (with no extra cost), and compute the local gradients in parallel for each model replica (using a small batch of data in the co-located Sample partition), as illustrated in Figure 3.

BigDL does not support model parallelism (i.e., no distribution of the model across different workers). This is not a limitation in practice, as BigDL runs on Intel Xeon CPU servers, which usually have large (100s of GB) memory size and can easily hold very large models.

\begin{figure*}
\centering
\includegraphics[width=1\linewidth]{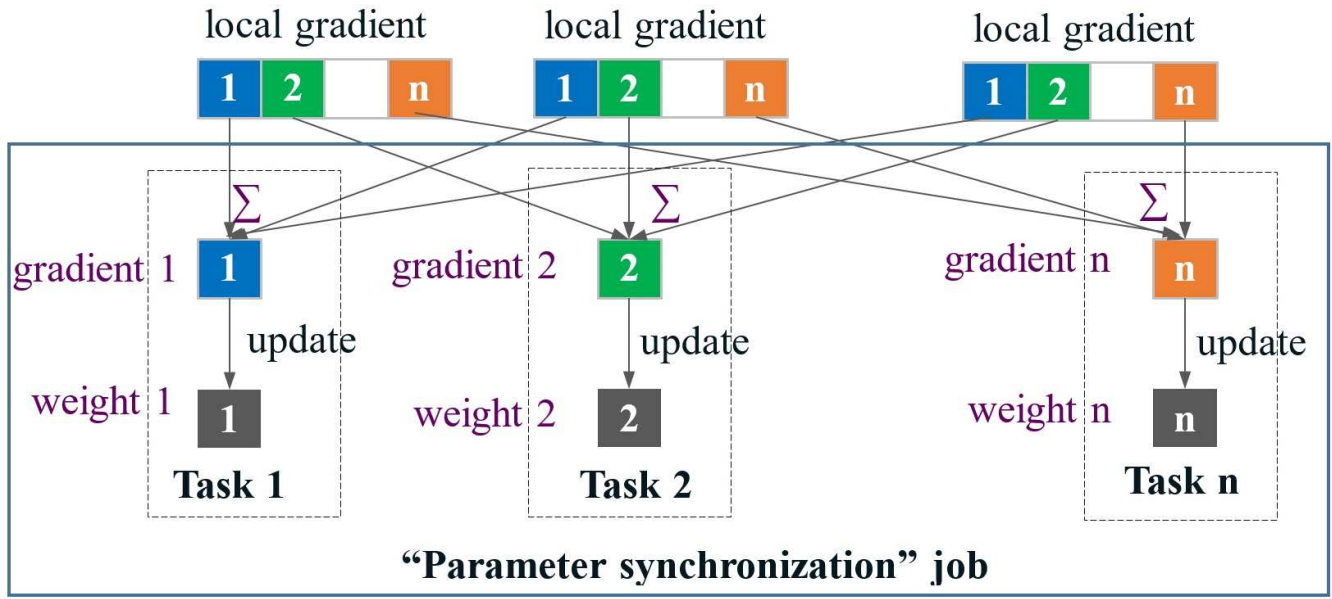}
\caption{Parameter synchronization in BigDL. Each local gradient (computed by a task in the "model forward-backward" job) is evenly divided into N partitions; then each task n in the "parameter synchronization" job aggregates these local gradients and updates the weights for the nth partition.}
\label{fig:fig4}
\end{figure*}

\subsection{Parameter synchronization in BigDL}

Parameter synchronization is a performance critical operation for data parallel distributed model training (in terms of speed and scalability). To support efficient parameter synchronization, existing deep learning frameworks usually implement parameter server or AllReduce using operations like fine-grained data access and in-place data mutation. Unfortunately, these operations are not supported by the functional compute model of big data systems (such as Spark).

\begin{table}[ht]
\centering
\begin{tabular}{l}
\toprule
\emph{  \emph{\textbf{Algorithm 2} "Parameter synchronization" job}} \\
\midrule
1: \textbf{for} each task \emph{n} in the "parameter synchronization" job \textbf{do} \\
2:\qquad \textbf{shuffle} the \emph{n$^{th}$} partition of all gradients to this task; \\
3:\qquad aggregate (sum) these gradients; \\
4:\qquad updates the \emph{n$^{th}$} partition of the weights; \\
5:\qquad \textbf{broadcast} the \emph{n$^{th}$} partition of the updated weights; \\
6: \textbf{end for} \\
\bottomrule
\end{tabular}
\end{table}

BigDL has taken a completely different approach that directly implements an efficient AllReduce like operation using existing primitives in Spark (e.g., shuffle, broadcast, in-memory cache, etc.), so as to mimic the functionality of a parameter server architecture (as illustrated in Figure 4).

\begin{itemize}
\item A Spark job has \emph{N} tasks, each of which is assigned a unique Id ranging from 1 to \emph{N} in BigDL. After each task in the "model forward-backward" job computes the local gradients (as described in Section 3.2 and illustrated in Figure 3), it evenly divides the local gradients into \emph{N} partitions, as shown in Figure 4.
\item Next, another "parameter synchronization" job is launched; each task \emph{n} of this job is responsible for managing the \emph{n$^{th}$} partition of the parameters (as shown in Algorithm 2), just like a parameter server does. Specifically, the \emph{n$^{th}$} partition of the local gradients (computed by the previous "model forward-backward" job) are first shuffled to task n, which aggregates these gradients and applies the updates to the \emph{n$^{th}$} partition of the weights, as illustrated in Figure 4.
\item After that, each task \emph{n} in the "parameter synchronization" job broadcasts the \emph{n$^{th}$} partition of the updated weights; consequently, tasks in the "model forward-backward" job of the next iteration can read the latest value of all the weights before the next training step begins.
\item The shuffle and task-side broadcast operations described above are implemented on top of the distributed in-memory storage in Spark: the relevant tasks simply store the local gradients and updated weights in the in-memory storage, which can then be read remotely by the Spark tasks with extremely low latency.
\end{itemize}

The implementation of AllReduce in BigDL has similar performance characteristics compared to \emph{Ring AllReduce} from Baidu Research \cite{gibiansky2017bringing}. As described in \cite{gibiansky2017bringing}, the total amount of data transferred to and from every node is \emph{2K(N-1)/N} in Ring AllReduce (where \emph{N} is the number of nodes and \emph{K} is the total size of the parameters); similarly, in BigDL, the total amount of data transferred to and from every node is \emph{2K}. In addition, all the bandwidth of every node in the cluster are fully utilized in both BigDL and Ring AllReduce. As a result, BigDL can efficiently train large deep neural network across large (e.g., hundreds of servers) clusters, as shown in Section 4.3.

\subsection{Discussions}

While BigDL has followed the standard practice (such as data parallel training and AllReduce operations) for scalable training, its implementation is very different from existing deep learning frameworks. By adopting the state of practice of big data systems (i.e., coarse-grained functional compute model), BigDL provides a viable design alternative for distributed model training. This allows deep learning algorithms and big data analytics to be seamless integrated into a single unified data pipeline, and completely eliminates the impedance mismatch problem described in Section 2. Furthermore, this also makes it easy to handle failures, resource changes, task preemptions, etc., which are expected to be norm rather than exception in large-scale systems.

Existing distributed deep learning frameworks (e.g., TensorFlow, MXNet, Petuum \cite{xing2015petuum}, ChainerMN \cite{akiba2017chainermn}, etc.) have adopted an architecture where multiple long-running, stateful tasks interact with others for model computation and parameter synchronization, usually in a blocking fashion to support synchronous distributed training. While this is optimized for constant communications among the tasks, it can only support coarse-grained failure recovery by completely starting over from previous (e.g., a couple of epochs before) snapshots.

In contrast, BigDL runs a series of short-lived Spark jobs (e.g., two jobs per mini-batch as described earlier), and each task in the job is stateless, non-blocking, and completely independent of each other; as a result, BigDL tasks can simply run without gang scheduling. In addition, it can also efficiently support fine-grained failure recovery by just re-running the failed task (which can then re-generate the associated partition of the local gradient or updated weight in the in-memory storage of Spark); this allows the framework to automatically and efficiently address failures (e.g., cluster scale-down, task preemption, random bugs in the code, etc.) in a timely fashion.

While AllReduce has been implemented in almost all existing deep learning frameworks, the implementation in BigDL is very different. In particular, existing deep learning frameworks usually implement the AllReduce operation using MPI-like primitives; as a result, they often create long-running task replicas that coordinate among themselves with no central control. On the other hand, BigDL has adopted a logically centralized control for distributed training \cite{liang2018rllib}; that is, a single driver program coordinates the distributed training (as illustrated in Algorithm 1). The driver program first launches the "model forward-backward" job to compute the local gradients, and then launches the "parameter synchronization" job to update the weights. The dependence between the two jobs are explicitly managed by the driver program, and each individual task in the two jobs are completely stateless and non-blocking once they are launched by the driver.

\section{Evaluation}

This section evaluates the computing performance and scalability of neural network training in BigDL. In addition, while we do not report inference performance results in this section, Section 5.1 shows the comparison of a real-world object detection inference pipeline running on BigDL vs. Caffe (and as reported by JD.com, the BigDL inference pipeline running on ~24 Intel Xeon servers is ~3.83x faster than Caffe running on 5 servers and 20 GPU cards ).

\subsection{Experiments}

Two categories of neural network models are used in this section to evaluate the performance and scalability of BigDL, namely, neural collaborative filtering (NCF) and convolutional neural network (CNN), which are representatives of the workloads that BigDL users run in their production Big Data platform.

Neural Collaborative Filtering (NCF) \cite{he2017neural} is one of most commonly used neural network models for recommendation, and has also been included in MLPerf  \cite{MLPerf}, a widely used benchmark suite for measuring training and inference performance of machine learning hardware, software, and services. In our experiments, we compare the training performance of BigDL (running on Intel Xeon server) vs. PyTorch (running on GPU).

In addition, deep convolutional neural networks (CNNs) have achieved human-level accuracy and are widely used for many computer vision tasks (such as image classifications and object detection). In our experiments, we study the scalability and efficiency of training Inception-v1 \cite{szegedy2015going} on ImageNet dataset \cite{deng2009imagenet} in BigDL with various number of Intel Xeon servers and Spark task; the results for other deep convolutional models, such as Inception-v3 \cite{szegedy2016rethinking} and ResNet50 \cite{he2016deep}, are similar. We do not include results for RNN (recurrent neural networks) training in this section, because it actually has better scalability compared to CNN training. This is because RNN computation is much slower than CNN, and therefore the parameter synchronization overhead (as a fraction of model compute time) is also much lower.

\subsection{Computing Performance}
To study the computing performance of BigDL, we compare the training speed of the NCF model using BigDL and PyTorch. MLPerf has provided a reference implementation of the NCF program \cite{PyTorch} based on PyTorch 0.4, which trains a movie recommender using the MovieLens 20Million dataset (ml-20m)  \cite{harper2016movielens}, a widely used benchmark dataset with 20 million ratings and 465,000 tags applied to 27,000 movies by 138,000 users. It also provides the reference training speed of the PyTorch implementation (to achieve the target accuracy goal) on a single Nvidia P100 GPU.

We have implemented the same NCF program using BigDL 0.7.0 and Spark 2.1.0 \cite{NCF}. We then trained the program on a dual-socket Intel Skylake 8180 2.5GHz server (with 56 cores in total and 384GB memory), and it took 29.8 minutes to converge and achieve the same accuracy goal.

\begin{figure}[ht]
\centering
\includegraphics[width=1\linewidth]{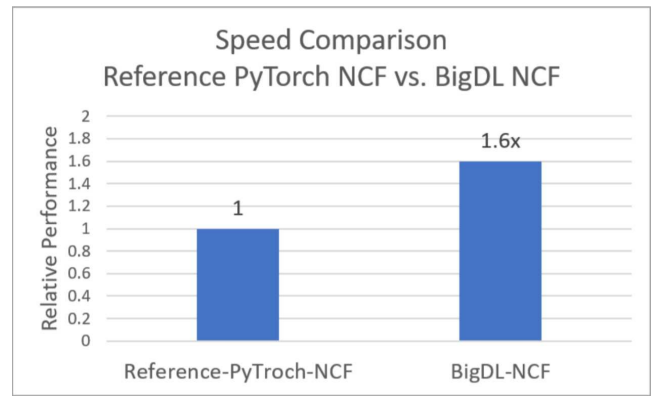}
\caption{The training performance of NCF using the BigDL implementation is 1.6x faster than the reference PyTorch implementation, as reported by MLPerf \cite{MLPerf05}.}
\label{fig:fig5}
\end{figure}

As reported by MLPerf, the training performance of NCF using the BigDL implementation is 1.6x faster than the reference PyTorch implementation \cite{MLPerf05} (as shown in Figure 5). While this only compares the training performance of BigDL on a single CPU server to PyTorch on a single GPU, it shows BigDL provides efficient implementations for neural network model computation (forward and backward). We will study the scalability and efficiency of the distributed training in BigDL in Section 4.3 and 4.4.

\subsection{Scalability of distributed training}

In the machine learning community, it is commonly believed that fine-grained data access and in-place data mutation are critical for efficient distributed training, and mechanisms like Spark's RDDs would impose significant overheads \cite{abadi2016tensorflow}. In this section, we show that BigDL provides highly efficient and scalable training, despite it is built on top of the coarse-grained functional compute model and immutable RDDs of Spark.

The scalability of distributed training in BigDL is determined by the efficiency (or overheads) of its parameter synchronizations. We first study the parameter synchronization overheads in BigDL by running ImageNet Inception-v1 model training using BigDL on various number of Xeon servers (dual-socket Intel Broadwell 2.20GHz, 256GB RAM and 10GbE network) \cite{BigDL2017}. As shown in Figure 6, the parameter synchronization overheads, measured as a fraction of the average model computation (forward and backward) time, turn out to be small (e.g., less than 7\% for Inception-v1 training on 32 nodes) in BigDL.

To study the scalability of the distributed training of BigDL on very large-scale Intel Xeon clusters, Cray have run ImageNet Inception-v1 model training using BigDL 0.3.0 with various node counts (starting at 16 nodes and scaling up to 256 nodes) \cite{Urika-XC}. Each node is a dual-socket Intel Broadwell 2.1 GHz (CCU 36 and DDR4 2400) server; the learning rate and Spark's executor memory are set to 0.10 and 120 GB respectively in the experiments.

\begin{figure}[ht]
\centering
\includegraphics[width=1\linewidth]{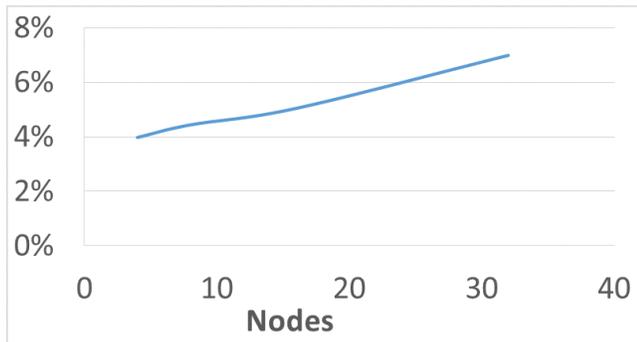}
\caption{Overheads of parameter synchronization (as a fraction of average model computation time) of ImageNet Inception-v1 training in BigDL \cite{BigDL2017}.}
\label{fig:fig6}
\end{figure}

\begin{figure}[ht]
\centering
\includegraphics[width=1\linewidth]{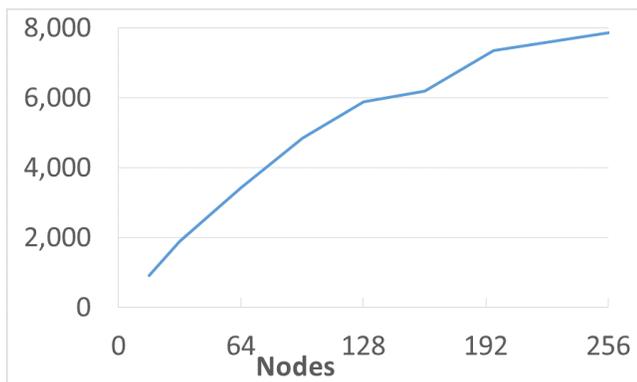}
\caption{Throughput of ImageNet Inception-v1 training in BigDL 0.3.0 reported by Cray, which scales almost linearly up to 96 nodes (and continue to scale reasonably up to 256 nodes) \cite{Urika-XC}.}
\label{fig:fig7}
\end{figure}

\begin{figure}[ht]
\centering
\includegraphics[width=1\linewidth]{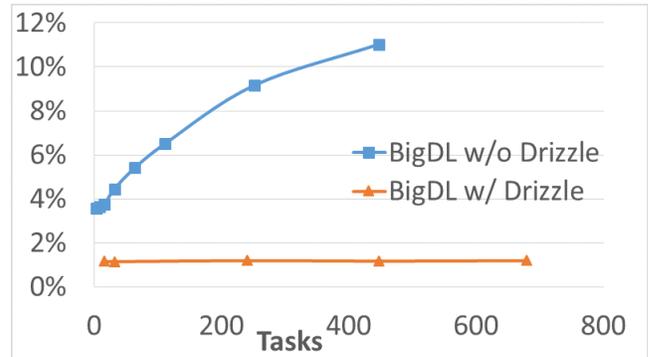}
\caption{Overheads of task scheduling and dispatch (as a fraction of average computation time) for ImageNet Inception-v1 training in BigDL  \cite{ApacheSpark}.}
\label{fig:fig8}
\end{figure}

Figure 7 shows the throughput of ImageNet Inception-v1 training; the training throughput scales almost linearly up to 96 nodes (e.g., about 5.3x speedup on 96 nodes compared to 16 nodes), and continue to scale reasonably well up to 256 nodes \cite{Urika-XC}. The results show that, even though BigDL implements its parameter server architecture directly on top of Spark (with immutable RDDs and coarse-grained functional operations), it can still provide efficient distributed training on large clusters.

\subsection{Efficiency of task scheduling}

As described in Section 3.4, BigDL needs to run a very large number of shot-lived tasks on Spark (e.g., the ImageNet Inception-v1 training may run 100s of thousands of iterations and 100s of tasks in parallel per iteration, while each task runs for just a couple of seconds); as a result, the underlying Spark framework needs to schedule a very large number of tasks across the cluster in a short period of time, which can potentially become a bottleneck on large clusters. For instance, Figure 8 shows that the overhead of launching tasks (as a fraction of average model computation time) in ImageNet Inception-v1 training on BigDL, while low for 100-200 tasks per iteration, can grows to over 10\% when there are close to 500 tasks per iteration \cite{ApacheSpark}.

To address this issue, in each training iteration BigDL will launch only a single (multi-threaded) task on each server, so as to achieve high scalability on large clusters (e.g., up to 256 machines, as described in Section 4.3). To scale to an even larger number (e.g., over 500) of servers, one can potentially leverage the iterative nature of model training (in which the same operations are executed repeatedly). For instance, group scheduling introduced by Drizzle \cite{Drizzle}, a low latency execution engine for Spark, can help schedule multiple iterations (or a group) of computations at once, so as to greatly reduce scheduling overheads even if there are a large number of tasks in each iteration, as shown in Figure 8 (which ran on AWS EC2 using r4.x2large instances) \cite{ApacheSpark}.

\begin{figure*}
\centering
\includegraphics[width=1\linewidth]{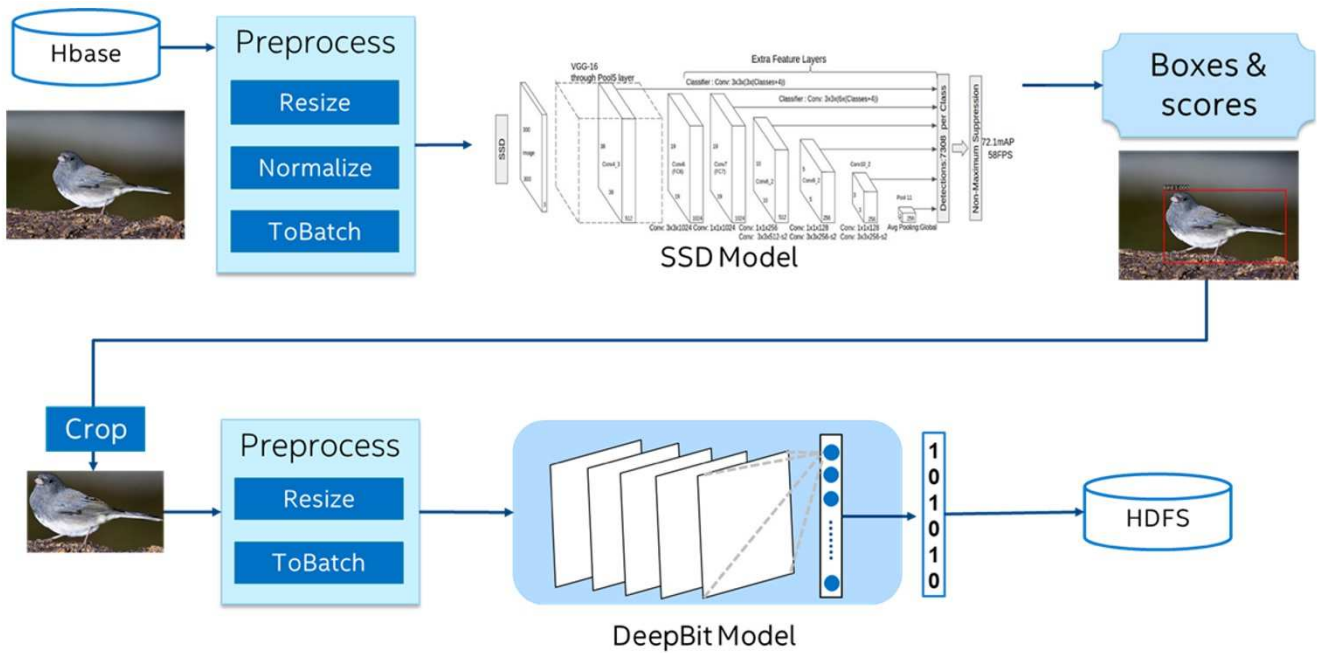}
\caption{End-to-end object detection and image feature extraction pipeline (using SSD and DeepBit models) on top of Spark and BigDL \cite{JD}.}
\label{fig:fig9}
\end{figure*}

\section{Applications}

Since its initial open source release (on Dec 30, 2016), BigDL users have built many deep learning applications on Spark and big data platforms. In this section, we share the real-world experience and "war stories" of our users that adopts BigDL to build the end-to-end data analysis and deep learning pipelines for their production data.

\begin{figure}[ht]
\centering
\includegraphics[width=1\linewidth]{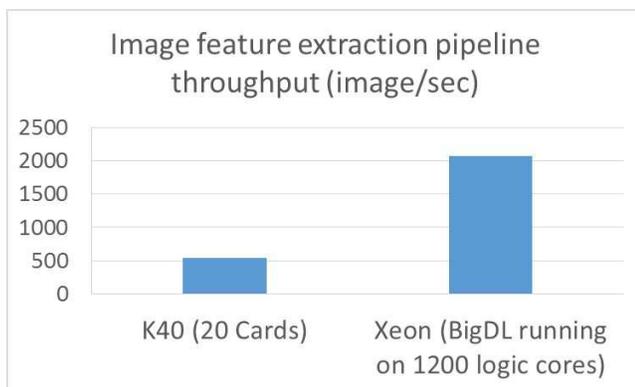}
\caption{Throughput of GPU clusters and Xeon clusters for the image feature extraction pipeline benchmarked by JD; the GPU cluster consists of 20 NVIDIA Tesla K40 cards, and the Xeon cluster consists of 1200 logical cores (with each Intel Xeon E5-2650 v4 2.2GHz server running 50 logical cores) \cite{JD}.}
\label{fig:fig10}
\end{figure}

\begin{figure*}
\centering
\includegraphics[width=1\linewidth]{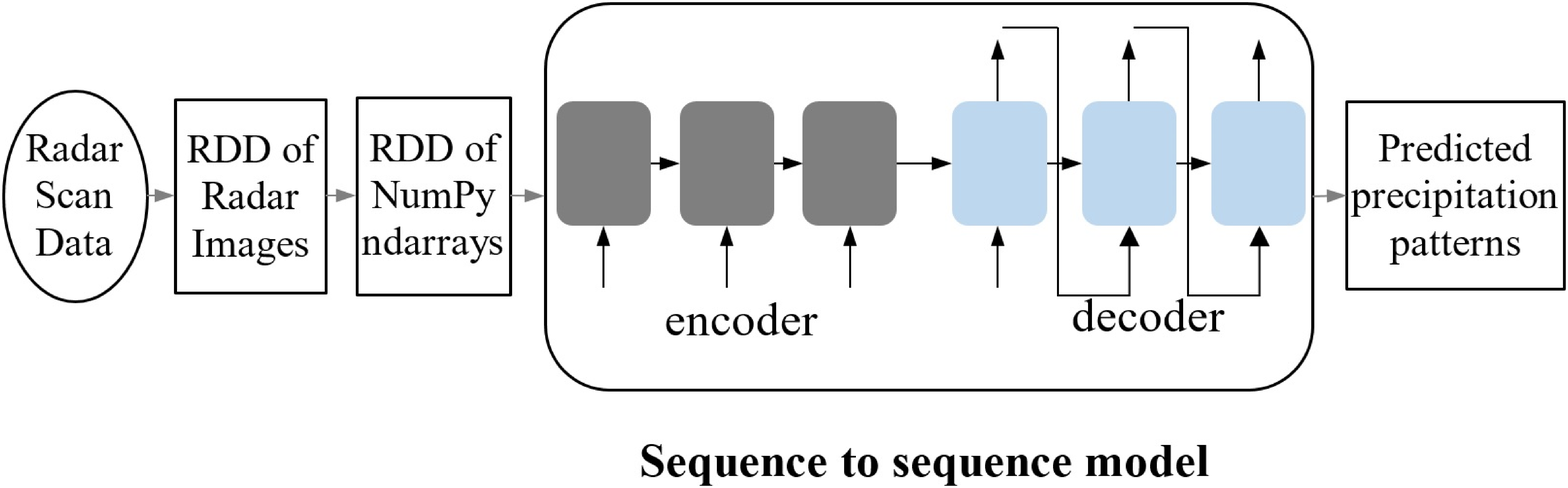}
\caption{End-to-end precipitation nowcasting workflow (using sequence-to-sequence models) on Spark and BigDL \cite{Urika-XC}.}
\label{fig:fig11}
\end{figure*}

\begin{figure*}
\centering
\includegraphics[width=0.85\linewidth]{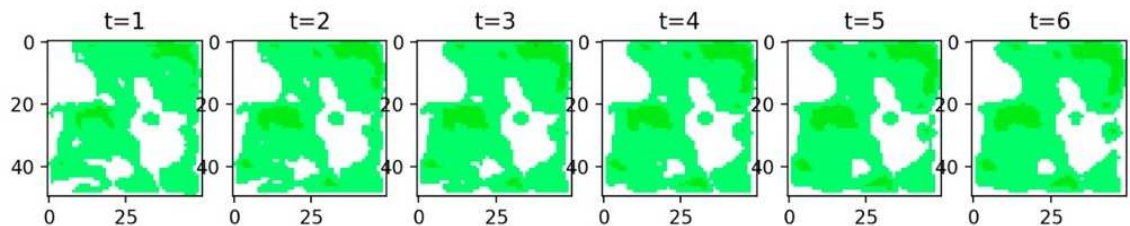}
\caption{Predicting precipitation patterns for the next hour (i.e., a sequence of images for the future time steps of the next hour) on Spark and BigDL \cite{Urika-XC}.}
\label{fig:fig12}
\end{figure*}

\subsection{Image feature extraction using object detection models}

JD.com has built an end-to-end object detection and image feature extraction pipeline on top of Spark and BigDL \cite{JD}, as illustrated in Figure 9.
\begin{itemize}
\item The pipeline first reads hundreds of millions of pictures from a distributed database into Spark (as an RDD of pictures), and then pre-processes the RDD of pictures in a distributed fashion using Spark.
\item It then uses BigDL to load a SSD \cite{ECCV2016} model (pre-trained in Caffe) for large scale, distributed object detection on Spark, which generates the coordinates and scores for the detected objects in each of the pictures.
\item It then generates the RDD of target images (by keeping the object with highest score as the target, and cropping the original picture based on the coordinates of the target), and further pre-processes the RDD of target images.
\item Finally it uses BigDL to load a DeepBit \cite{IEEE2016} model (again pre-trained in Caffe) for distributed feature extraction of the target images, and stores the results (RDD of extracted object features) in HDFS.
\end{itemize}
Previously JD engineers have deployed the same solution on a 5-node GPU cluster with 20 NVIDIA Tesla K40 following a "connector approach" (similar to CaffeOnSpark): reading data from HBase, partitioning and processing the data across the cluster, and then running the deep learning models on Caffe. This turns out to be very complex and error-prone (because all of the data partitioning, load balancing, fault tolerance, etc., need to be manually managed). In addition, it also reveals an impedance mismatch of the "connector approach" (HBase + Caffe in this case) - reading data from HBase takes about half of the time in this solution (because the task parallelism is tied to the number of GPU cards in the system, which is too low for interacting with HBase to read the data).

After migrating the solution to BigDL, JD engineers can easily implement the entire data analysis and deep learning pipeline (including data loading, partitioning, pre-processing, model inference, etc.,) under a unified programming paradigm on Spark. This not only greatly improves the efficiency of development and deployment, but also delivers about 3.83x speedup (running on about 24 Intel Broadwell 2.2GHz servers) compared to running the Caffe-based solution on the GPU cluster (with 20 NVIDIA Tesla K40 cards), as reported by JD \cite{JD} and shown in Figure 10.

\subsection{Precipitation nowcasting using Seq2Seq models}

Cray has built a precipitation nowcasting (predicting short-term precipitation) application using a Seq2Seq \cite{seq2seq} model (with a stacked convolutional LSTM network \cite{lstm} as the encoder, and another stacked convolutional LSTM network as the decoder); the end-to-end pipeline runs on Spark and BigDL \cite{Urika-XC}, including data preparation, model training and inference (as illustrated in Figure 11).

\begin{itemize}
\item The application first reads over a terabyte of raw radar scan data into Spark (as an RDD of radar images), and then converts it into an RDD of NumPy ndarrays.
\item It then trains a sequence-to-sequence model, using a sequence of images leading up to the current time as the input, and a sequence of predicted images in the future as the output.
\item After the model is trained, it can be used to predict, say, the precipitation patterns (i.e., a sequence of images for the future time steps) of the next hour, as illustrated in Figure 12.
\end{itemize}

Cray engineers have previously implemented the application using two separate workflows: running data processing on a highly distributed Spark cluster, and deep learning training on another GPU cluster running TensorFlow. It turns out that this approach not only brings high data movement overheads, but also greatly hurts the development productivity due to the fragmented workflow. As a result, Cray engineers chose to implement the solution using a single unified data analysis and deep learning pipeline on Spark and BigDL, which greatly improves the efficiency of development and deployment.

\subsection{Real-time streaming speech classification}

\begin{figure*}
\centering
\includegraphics[width=0.9\linewidth]{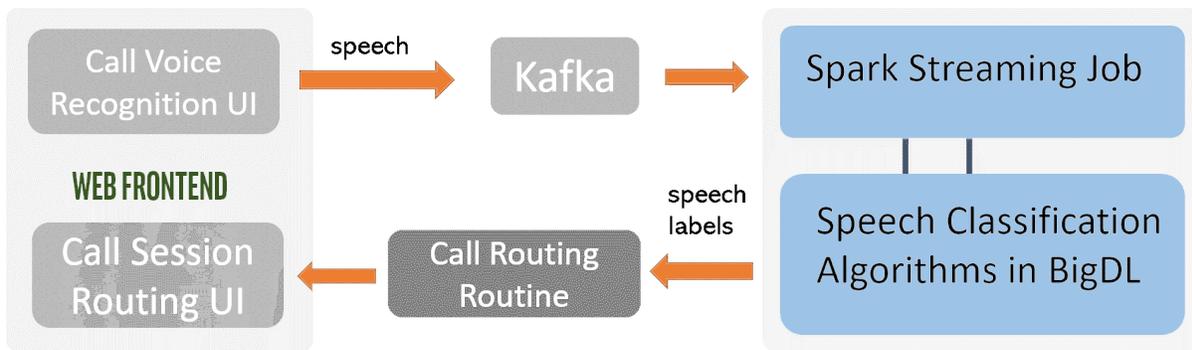}
\caption{The end-to-end workflow of real-time streaming speech classification on Kafka, Spark Streaming and BigDL \cite{GigaSpaces}.}
\label{fig:fig13}
\end{figure*}

GigaSpaces has built a speech classification application for efficient call center management \cite{GigaSpaces}, which automatically routes client calls to corresponding support specialists in real-time. The end-to-end workflow is implemented using BigDL with Apache Kafka \cite{Kafka} and Spark Streaming \cite{streams} (as illustrated Figure 13), so as to provide distributed realtime streaming model inference.
\begin{itemize}
\item When a customer calls the call center, his or her speech is first processed on the fly by a speech recognition unit and result is stored in Kafka.
\item A Spark Streaming job then reads speech recognition results from Kafka and classifies each call using the BigDL model in real-time.
\item The classification result is in turn used by a routing system to redirect the call to the proper support specialist.
\end{itemize}

One of the key challenges for GigaSpaces engineers to implement the end-to-end workflow is how to efficiently integrate the new neural network models in the realtime stream processing pipeline, and how to seamlessly scale the streaming applications from a handful machines to thousands of nodes. BigDL allows neural network models to be directly applied in standard distributed streaming architecture for Big Data (using Apache Kafka and Spark Streaming), which can then efficiently scales out to a large number of nodes in a transparent fashion. As a result, this greatly improves the developer productivity and deployment efficiency of the end-to-end streaming workflow.

\section{Related Work}

Existing deep learning frameworks (such as TensorFlow, MXNet, Petuum, ChainerMN, etc.) typically provide efficient parameter server and/or AllReduce implementation (using fine-grained data access and in-place data mutation) for distributed training. In contrast, BigDL provides distributed training support directly on top of a functional compute model of big data systems (with copy-on-write and coarse-grained operations), which is completely different from the implementation in existing deep learning frameworks. This provides a viable design alternative for distributed model training by adopting the state of practice of big data systems, and makes it easy to handle failures, resource changes, task preemptions, etc., in a more timely and fine-grained fashion.

As discussed in Section 2, to address the challenge of integrating deep learning into real-world data pipelines, there have been many efforts in the industry that adopt a "connector approach" (e.g., TFX, CaffeOnSpark, TensorFlowOnSpark, SageMaker, etc.). Unfortunately, these frameworks can incur very large overheads in practice due to the adaptation layer between different frameworks; more importantly, they often suffer from impedance mismatches that arise from crossing boundaries between heterogeneous components. While efforts in the Big Data community (such as Project Hydrogen in Spark) attempt to overcome some of these issues brought by the "connector approach", they still do not address the fundamental "impedance mismatch" problem (as discussed in Section 2). By unifying the distributed execution model of deep neural network models and big data analysis, BigDL provides a single unified data pipeline for both deep learning and big data analysis, which eliminates the adaptation overheads or impedance mismatch.

\section{Summary}

We have described BigDL, including its distributed execution model, computation performance, training scalability, and real-world use cases. It allows users to build deep learning applications for big data using a single unified data pipeline; the entire pipeline can directly run on top of existing big data systems in a distributed fashion. Unlike existing deep learning frameworks, it provides efficient and scalable distributed training directly on top of the functional compute model (with copy-on-write and coarse-grained operations) of Spark. BigDL is a work in progress, but our initial experience is encouraging. Since its initial open source release on Dec 30, 2016, it has received over 3100 stars on Github; and it has enabled many users (e.g., Mastercard, World Bank, Cray, Talroo, UCSF, JD, UnionPay, Telefonica, GigaSpaces, etc.) to build new analytics and deep learning applications for their production data pipelines.

%%
%% The next two lines define the bibliography style to be used, and
%% the bibliography file.
\bibliographystyle{ACM-Reference-Format}
\bibliography{socc2019-sample-base}

\begin{thebibliography}{10}

\bibitem{jia2014caffe}
Jia, Yangqing and Shelhamer, Evan and Donahue, Jeff and Karayev, Sergey and
  Long, Jonathan and Girshick, Ross and Guadarrama, Sergio and Darrell, Trevor.
  Caffe: Convolutional architecture for fast feature embedding. in {\em
  Proceedings of the 22nd ACM international conference on Multimedia}.
\newblock MM'14.

\bibitem{collobert2011torch7}
Collobert, Ronan and Kavukcuoglu, Koray and Farabet, Cl{\'e}ment. Torch7: A
  matlab-like environment for machine learning. in {\em BigLearn, NIPS
  workshop}.
\newblock (2011).

\bibitem{abadi2016tensorflow}
Abadi, M., Barham, P., Chen, J., Chen, Z., Davis, A., Dean, J., Devin, M.,
  Ghemawat, S., Irving, G., Isard, M., Kudlur, M., Levenberg, J., Monga, R.,
  Moore, S., Murray, D. G., Steiner, B., Tucker, P., Vasudevan, V., Warden, P.,
  Wicke, M., Yu, Y., and Zheng, X. Tensorflow: A system for large-scale machine
  learning. in {\em Proceedings of the 12th USENIX Conference on Operating
  Systems Design and Implementation.}
\newblock OSDI'16.

\bibitem{chen2015mxnet}
Chen, Tianqi and Li, Mu and Li, Yutian and Lin, Min and Wang, Naiyan and Wang,
  Minjie and Xiao, Tianjun and Xu, Bing and Zhang, Chiyuan and Zhang, Zheng.
  Mxnet: A flexible and efficient machine learning library for heterogeneous
  distributed systems.
\newblock {\em In Proceedings of Workshop on Machine Learning Systems
  (LearningSys) in The Twenty-ninth Annual Conference on Neural Information
  Processing Systems (NIPS)}.
\newblock (2015).

\bibitem{tokui2015chainer}
Tokui, Seiya and Oono, Kenta and Hido, Shohei and Clayton, Justin Chainer: a
  next-generation open source framework for deep learning in {\em In
  Proceedings of workshop on machine learning systems (LearningSys) in the
  twenty-ninth annual conference on neural information processing systems
  (NIPS)}.
\newblock (2015).

\bibitem{paszke2017automatic}
Paszke, Adam and Gross, Sam and Chintala, Soumith and Chanan, Gregory and Yang,
  Edward and DeVito, Zachary and Lin, Zeming and Desmaison, Alban and Antiga,
  Luca and Lerer, Adam Automatic differentiation in pytorch.
\newblock {\em NIPS 2017 Autodiff Workshop}.
\newblock (2017).

\bibitem{spark}
Apache spark Apache software foundation.
\newblock (2014) (https://spark.apache.org).

\bibitem{hadoop}
Apache hadoop Apache software foundation.
\newblock (2006) (https://hadoop.apache.org).

\bibitem{keras}
Chollet,F.et al. Keras.
\newblock (https://keras.io).

\bibitem{zaharia2012resilient}
Zaharia, Matei and Chowdhury, Mosharaf and Das, Tathagata and Dave, Ankur and
  Ma, Justin and McCauley, Murphy and Franklin, Michael J and Shenker, Scott
  and Stoica, Ion. Resilient distributed datasets: A fault-tolerant abstraction
  for in-memory cluster computing. in {\em Proceedings of the 9th USENIX
  conference on Networked Systems Design and Implementation}.
\newblock NSDI'12.

\bibitem{armbrust2015spark}
Armbrust, Michael and Xin, Reynold S and Lian, Cheng and Huai, Yin and Liu,
  Davies and Bradley, Joseph K and Meng, Xiangrui and Kaftan, Tomer and
  Franklin, Michael J and Ghodsi, Ali and others. Spark sql: Relational data
  processing in spark. in {\em 2015 ACM SIGMOD international conference on
  management of data.}
\newblock SIGMOD'15.

\bibitem{russakovsky2015imagenet}
Russakovsky, Olga and Deng, Jia and Su, Hao and Krause, Jonathan and Satheesh,
  Sanjeev and Ma, Sean and Huang, Zhiheng and Karpathy, Andrej and Khosla,
  Aditya and Bernstein, Michael and others. Imagenet large scale visual
  recognition challenge.
\newblock {\em International journal of computer vision(IJCV)}.
\newblock (2015).

\bibitem{rajpurkar2016squad}
Rajpurkar,P and Zhang,J and Lopyrev,K and Liang,P. Squad: 100,000+ questions
  for machine comprehension of text.
\newblock {\em In Proceedings of the 2016 Conference on Empirical Methods in
  Natural Language Processing, EMNLP.}
\newblock (2016).

\bibitem{jawaheer2010comparison}
Jawaheer,G and Szomszor,M and Kostkova,P. Comparison of implicit and explicit
  feedback from an online music recommendation service. in {\em proceedings of
  the 1st international workshop on information heterogeneity and fusion in
  recommender systems.}
\newblock (2010) HetRec'10.

\bibitem{baylor2017tfx}
Baylor, D., Breck, E., Cheng, H.-T., Fiedel, N., Foo, C. Y., Haque, Z., Haykal,
  S., Ispir, M., Jain, V., Koc, L., Koo, C. Y., Lew, L., Mewald, C., Modi, A.
  N., Polyzotis, N., Ramesh, S., Roy, S., Whang, S. E., Wicke, M., Wilkiewicz,
  J., Zhang, X., and Zinkevich, M. Tfx: A tensorflow-based production-scale
  machine learning platform in {\em Proceedings of the 23rd ACM SIGKDD
  International Conference on Knowledge Discovery and Data Mining}.
\newblock KDD'17.

\bibitem{Yahoo2016}
CaffeOnSpark. Yahoo.
\newblock (2016) (https://github.com/yahoo/CaffeOnSpark).

\bibitem{Yahoo2017}
TensorflowOnSpark. Yahoo.
\newblock (2017) (https://github.com/yahoo/TensorFlowOnSpark).

\bibitem{Amazon2017}
Sagemaker. Amazon.
\newblock (2017) (https://aws.amazon.com/sagemaker/).

\bibitem{lin2013scaling}
Lin, Jimmy and Ryaboy, Dmitriy Scaling big data mining infrastructure: the
  twitter experience.
\newblock {\em ACM SIGKDD Explorations Newsletter} 14(2).
\newblock (December 2012).

\bibitem{ReynoldXin2018}
Reynold Xin. "project hydrogen: Unifying state-of-the-art ai and big data in
  apache spark". spark + ai summit 2018.

\bibitem{GangScheduling}
Gang scheduling.
\newblock (https://en.wikipedia.org/wiki/Gang\_scheduling/).

\bibitem{zaharia2010delay}
Zaharia, Matei and Borthakur, Dhruba and Sen Sarma, Joydeep and Elmeleegy,
  Khaled and Shenker, Scott and Stoica, Ion. Delay scheduling: a simple
  technique for achieving locality and fairness in cluster scheduling. in {\em
  Proceedings of the 5th European conference on Computer systems,}.
\newblock EuroSys'10.

\bibitem{dean2012large}
Dean,J., Corrado,G., Monga,R., Chen,K., Devin,M., Mao,M., Ranzato,Marc'aurelio,
  Senior,A., Tucker,P., Yang,K., Le,Q.V., Ng,A.Y. Large scale distributed deep
  networks. in {\em Proceedings of the 25th International Conference on Neural
  Information Processing Systems}.
\newblock NIPS'12.

\bibitem{li2014scaling}
Li,M., Andersen,D.G., Park,J.W., Smola,A.J., Ahmed,A., Josifovski,V., Long,J.,
  Shekita,E.J., and Su,B.-Y. Scaling distributed machine learning with the
  parameter server. in {\em Proceedings of the 11th USENIX Conference on
  Operating Systems Design and Implementation}.
\newblock OSDI'14.

\bibitem{chilimbi2014project}
Chilimbi,T., Suzue,Y., Apacible,J., and Kalyanaraman,K. Project adam: Building
  an efficient and scalable deep learning training system. in {\em Proceedings
  of the 11th USENIX Conference on Operating Systems Design and
  Implementation}.
\newblock OSDI'14.

\bibitem{xing2015petuum}
Xing,E.P., Ho,Q., Dai,W., Kim,J.-K., Wei,J., Lee,S., Zheng,X., Xie,P.,
  Kumar,A., and Yu,Y. Petuum: A new platform for distributed machine learning
  on big data.
\newblock {\em Proceedings of the 21th ACM SIGKDD International Conference on
  Knowledge Discovery and Data Mining}.
\newblock KDD'15.

\bibitem{zhang2017poseidon}
Zhang,H., Zheng,Z., Xu,S., Dai,W., Ho,Q., Liang,X., Hu,Z., Wei,J., Xie,P., and
  Xing,E.P. Poseidon: An efficient communication architecture for distributed
  deep learning on gpu clusters. in {\em 2017 USENIX Annual Technical
  Conference (USENIX ATC 17)}.
\newblock (2017).

\bibitem{dean2008mapreduce}
Jeffrey Dean, Sanjay Ghemawat Mapreduce: simplified data processing on large
  clusters.
\newblock {\em Proceedings of the 6th conference on Symposium on Operating
  Systems Design \& Implementation,$\{$OSDI$\}$}.
\newblock (2004).

\bibitem{isard2007dryad}
Michael Isard, Mihai Budiu, Yuan Yu, Andrew Birrell, and Dennis Fetterly.
  Dryad: distributed data-parallel programs from sequential building blocks in
  {\em Proceedings of the 2nd ACM SIGOPS/EuroSys European Conference on
  Computer Systems 2007}.
\newblock EuroSys'07.

\bibitem{chen2016revisiting}
Chen,J., Monga,R., Bengio,S., and Jozefowicz,R. Revisiting distributed
  synchronous sgd.
\newblock {\em In International Conference on Learning Representations Workshop
  Track.}
\newblock (2016).

\bibitem{gibiansky2017bringing}
Gibiansky,Andrew. "bringing hpc techniques to deep learning".
\newblock (http://andrew.gibiansky.com/blog/machine-learning/baidu-allreduce/).

\bibitem{akiba2017chainermn}
Akiba,T., Fukuda,K., and Suzuki,S. Chainermn: Scalable distributed deep
  learning framework.
\newblock {\em Proceedings of Workshop on ML Systems in The Thirty-first Annual
  Conference on Neural Information Processing Systems (NIPS)}.
\newblock (2017).

\bibitem{liang2018rllib}
Eric Liang, Richard Liaw, Philipp Moritz, Robert Nishihara, Roy Fox, Ken
  Goldberg, Joseph E. Gonzalez, Michael I. Jordan, Ion Stoica. Rllib:
  Abstractions for distributed reinforcement learning.
\newblock {\em International Conference on Machine Learning (ICML)}.
\newblock (2018).

\bibitem{he2017neural}
He, Xiangnan and Liao, Lizi and Zhang, Hanwang and Nie, Liqiang and Hu, Xia and
  Chua, Tat-Seng Neural collaborative filtering. in {\em Proceedings of the
  26th international conference on world wide web. International World Wide Web
  Conferences Steering Committee}.
\newblock (2017).

\bibitem{MLPerf}
Mlperf.
\newblock (https://mlperf.org/).

\bibitem{szegedy2015going}
Szegedy,C., Liu,W., Jia,Y., Sermanet,P., Reed,S., Anguelov,D., Erhan,D.,
  Vanhoucke,V., and Rabinovich,A. Going deeper with convolutions in {\em
  Computer Vision and Pattern Recognition (CVPR)}.
\newblock (2015).

\bibitem{deng2009imagenet}
Deng,J., Socher,R., Fei-Fei,L., Dong,W., Li,K., and Li,L.-J. Imagenet: A
  large-scale hierarchical image database. in {\em 2009 IEEE conference on
  computer vision and pattern recognition(CVPR)}.
\newblock (2009).

\bibitem{szegedy2016rethinking}
Szegedy,C., Vanhoucke,V., Ioffe,S., Shlens,J., and Wojna,Z. Rethinking the
  inception architecture for computer vision in {\em 2016 IEEE Conference on
  Computer Vision and Pattern Recognition (CVPR)}.
\newblock (2016).

\bibitem{he2016deep}
He, Kaiming and Zhang, Xiangyu and Ren, Shaoqing and Sun, Jian Deep residual
  learning for image recognition. in {\em Proceedings of the IEEE conference on
  computer vision and pattern recognition}.
\newblock (2016).

\bibitem{PyTorch}
Reference ncf implementation using pytorch in mlperf.
\newblock
  (https://github.com/mlperf/training/blob/\\master/recommendation/pytorch/README.md).

\bibitem{harper2016movielens}
Harper, F Maxwell and Konstan, Joseph A. "the movielens datasets: History and
  context".
\newblock {\em ACM Trans. Interact. Intell. Syst.} 5(4):19.
\newblock (2015).

\bibitem{NCF}
Ncf implementation in bigdl.
\newblock
  (https://github.com/mlperf/training\_results\_v0.5/tree\\/master/v0.5.0/intel/intel\_ncf\_submission).

\bibitem{MLPerf05}
Mlperf 0.5 training results.
\newblock (https://mlperf.org/training-results-0-5).

\bibitem{BigDL2017}
Jason (Jinquan) Dai, and Ding Ding. Very large-scale distributed deep learning
  with bigdl. o'reilly ai conference, san francisco.
\newblock (2017).

\bibitem{Urika-XC}
Alex Heye, et al. "scalable deep learning with bigdl on the urika-xc software
  suite".
\newblock
  (https://www.cray.com/blog/scalable-deep-learning-bigdl-urika-xc-software-suite/).

\bibitem{ApacheSpark}
Shivaram Venkataraman, et al. "accelerating deep learning training with bigdl
  and drizzle on apache spark".
\newblock
  (https://rise.cs.berkeley.edu/blog/accelerating-deep-learning-training-with-bigdl-and-drizzle-on-apache-spark/).

\bibitem{Drizzle}
Venkataraman,S., Panda,A., Ousterhout,K., Armbrust,M., Ghodsi,A.,
  Franklin,M.J., Recht,B., and Stoica,I. Drizzle: Fast and adaptable stream
  processing at scale in {\em Proceedings of the 26th Symposium on Operating
  Systems Principles}.
\newblock SOSP'17.

\bibitem{JD}
Jason (Jinquan) Dai, et al. Building large-scale image feature extraction with
  bigdl at jd.com.
\newblock
  (https://software.intel.com/en-us/articles/building-large-scale-image-feature-extraction-with-bigdl-at-jdcom).

\bibitem{ECCV2016}
Liu,W., Anguelov,D., Erhan,D., Szegedy,C., Reed,S.E., Fu,C.-Y., and Berg,A.C.
  Ssd: Single shot multibox detector in {\em ECCV}.
\newblock (2016).

\bibitem{IEEE2016}
Lin, K., Lu, J., Chen, C.-S., and Zhou, J. Learning compact binary descriptors
  with unsupervised deep neural networks. in {\em 2016 IEEE Conference on
  Computer Vision and Pattern Recognition (CVPR)}.
\newblock (2016).

\bibitem{seq2seq}
Sutskever,I., Vinyals,O., and Le,Q.V. Sequence to sequence learning with neural
  networks. in {\em Proceedings of the 27th International Conference on Neural
  Information Processing Systems}.
\newblock Vol.{}~2.
\newblock NIPS'14.

\bibitem{lstm}
Shi,X., Chen,Z., Wang,H., Yeung,D.-Y., Wong,W.-k., and Woo,W.-c. Convolutional
  lstm network: A machine learning approach for precipitation nowcasting. in
  {\em Proceedings of the 28th International Conference on Neural Information
  Processing Systems}.
\newblock Vol.{}~1.
\newblock NIPS'15.

\bibitem{GigaSpaces}
Rajiv Shah. Gigaspaces integrates insightedge platform with intel's bigdl for
  scalable deep learning innovation.
\newblock
  (https://www.gigaspaces.com/blog/gigaspaces-to-demo-with-intel-at-strata-data-conference-and-microsoft-ignite/).

\bibitem{Kafka}
Apache Kafka.
\newblock (https://kafka.apache.org/).

\bibitem{streams}
Matei Zaharia, Tathagata Das, Haoyuan Li, Timothy Hunter, Scott Shenker, and
  Ion Stoica. Discretized streams: fault-tolerant streaming computation at
  scale in {\em The Twenty-Fourth ACM Symposium on Operating Systems
  Principles}.
\newblock (2013) SOSP'13.

\end{thebibliography}

%%
%% If your work has an appendix, this is the place to put it.
\appendix

\end{document}